# Google as God?
# Opportunities and Risks of the Information Age

Dirk Helbing, ETH Zurich, Clausiusstrasse 50, 8092 Zurich, Switzerland
& ETH Risk Center, Scheuchzerstrasse 7, 8092 Zurich, Switzerland

***If God did not exist – people would invent one! The development of human civilization requires mechanisms promoting cooperation and social order. One of these mechanisms is based on the idea that everything we do is seen and judged by God – bad deeds will be punished, while good ones will be rewarded. The Information Age has now fueled the dream that God-like omniscience and omnipotence can be created by man.***

*1. Introduction*

> *"You're already a walking sensor platform… You are aware of the fact that somebody can know where you are at all times because you carry a mobile device, even if that mobile device is turned off. You know this, I hope? Yes? Well, you should… Since you can't connect dots you don't have, it drives us into a mode of, we fundamentally try to collect everything and hang on to it forever… It is really very nearly within our grasp to be able to compute on all human generated information."*
>
> *Ira "Gus" Hunt, CIA Chief Technology Officer* [1]

For many decades, the processing power of computer chips has increased exponentially – as predicted by "Moore's Law". Storage capacity is growing even faster. We are now entering a phase of the "Internet of Things", where computer chips and measurement sensors will soon be scattered everywhere producing huge amounts of data ("Big Data").  Cell phones, computers and factories, but also our coffee machines, fridges, shoes and clothes, among others are getting more and more connected.

*2. Gold Rush for the 21$^{st}$ Century Oil*

This huge amount of data, including credit card transactions, communication with friends and colleagues, mobility data and more is already celebrated as the "Oil of the 21st Century". The gold rush to exploit this valuable resource is just starting. The more data are generated, stored and interpreted, the easier is it for companies and secret services to know us better than our friends and families do. For example, the company "Recorded Future" – apparently a joint initiative between Google and the CIA – seems to investigate people's social networks and mobility profiles. Furthermore, credit card companies analyze "consumers' genes" – the factors that determine our consumer behavior.

---

[1] See: http://www.businessinsider.com/cia-presentation-on-big-data-2013-3?op=1 and http://gigaom.com/2013/03/20/even-the-cia-is-struggling-to-deal-with-the-volume-of-real-time-social-data/2/. For similar recent FBI priorities see http://www.slate.com/blogs/future_tense/2013/03/26/andrew_weissmann_fbi_wants_real_time_gmail_dropbox_spying_power.html

Our individual motivations are analyzed in order to understand our decisions and influence our behavior through personalized search, individualized advertisements, and recommendations or decisions of our Facebook friends. But how many of these "friends" are trustable, how many of them are paid to influence us, and how many are software robots?

*3. Humans Controlled by Computers?*

Today, computers autonomously perform the majority of financial transactions. They decide how much we have to pay for our loans or insurances, based on our behavioral data and on those of our friends, neighbors and colleagues. People are increasingly discriminated by obscure "machine learning" algorithms, which are neither transparent nor have to meet particular quality standards. People classified as dangerous are now eliminated by drones, without a chance to prove their innocence, while some countries are discussing robots rights. Soon, Google will drive our cars. And in ten years, supercomputers will exceed the performance of human brains.

*4. Is Privacy Still Needed?*

What will the role of privacy be in such an information society? Some companies are already trying to turn privacy into a marketable commodity. This is done by first taking away our privacy and then selling it back to us. The company Acxiom, for example, is said to sell detailed data about more than 500 million people. Would it be possible to know beforehand whether the data will be used for good or bad? Many will pay to have their personal data removed from the Internet and commercial databases. And where data removal is not possible, fake identities and mobility profiles will be offered for sale, to obfuscate our traces.

*5. Information Overload*

"Big Data" do not necessarily mean that we'll see the world more accurately. Rather, we will have to pay for "digital eyewear" that allows us to keep an overview in the data deluge. Those not willing to pay (possibly also with personal data) will be blinded by an information overload. Already today, we cannot assess the quality of the answers we get online. As the way in which the underlying data are processed remains hidden to the user, it is hard to know how much we are being manipulated by web services and social media. But given the huge economic potential, it is pretty clear that manipulation is happening.

*6. The Knowledge-Is-Power Society*

The statement "knowledge is power" seems to imply that "omniscience is omnipotence" – a tempting idea indeed. Therefore, who collects all the data in the world, such as the National Security Agency (NSA) in the United States, might hope to become almighty, especially if equipped with suitable manipulation tools. By knowing everything about us, one can always find an Achille's heel. Even CIA director General David Petraeus was not immune to this risk. He became the victim of a love affair irrelevant to his duty.

The developments outlined above are not fantasy - they are already taking place behind the scenes or are just around the corner. Yet, our society and legal system are not well prepared for this.

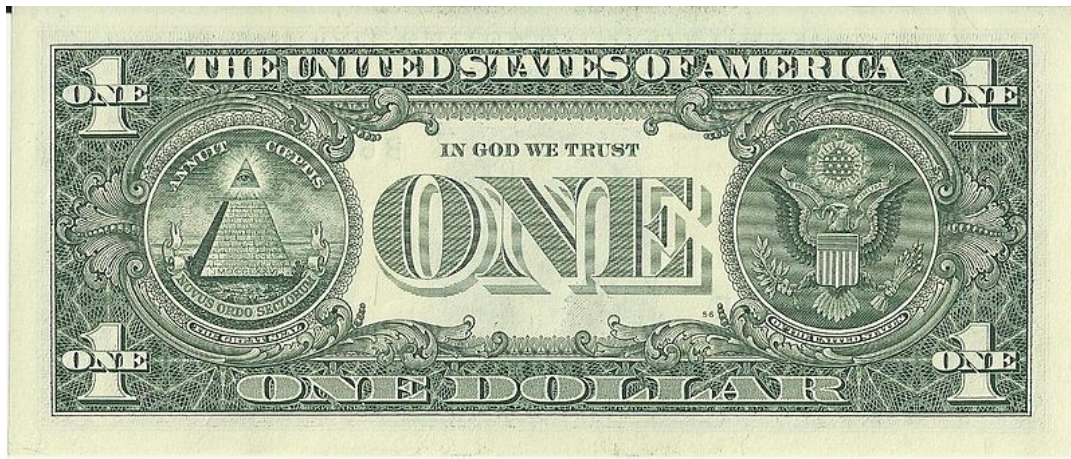

*Each 1 dollar bill suggests that omniscience and omnipotence (see the "God's eye") together with the belief in God ("In God We Trust") would be the basis of a new world order ("Novus Ordo Seclorum"). Source: Wikimedia Common, see*
http://en.wikipedia.org/wiki/File:US_one_dollar_bill,_reverse,_series_2009.jpg

*7. A New World Order Based on Information?*

Some people may see information and technologies as new tools to create social order. Why should one object to a computer or company or government taking decisions for us, as long as they act in our interest? But who would decide how to use these tools? Can the concept of a 'caring state' or a 'benevolent dictator' really work? In other words, can supercomputers enabled by Big Data take the best decisions for us?

This has always failed in the past, and will be unsuccessful in the future. Not only do many systems fail under asymmetric information (if some stakeholders are very well informed and others very badly). The performance of all computers in the world will never be sufficient to optimize our economy and society in real time. Supercomputers cannot even optimize the traffic lights of a big city in real time. This is because the computational effort explodes with the size and complexity of the system. Just a very simple society could be optimized top down, but who would want to live in it?

*8. Privacy and Socio-Diversity Need Protection*

The aforementioned "omniscient almighty society" cannot work. If we all did what is right according to a super-intelligent institution – it would be as if children would always do what their parents are asking for. Then they would never learn to take autonomous decisions, and to go their own way. Privacy is a necessary ingredient for the development of individual responsibility and for society. It should not to be understood as a concession to the citizens.

"Private" and "public" are two sides of the same coin, which could not exist without each other. People can only adjust to the thousands of normative public expectations every day, if there is a private, protected space where they can be free and relax. Privacy reduces mutual interference to a degree that allows us to "live and let live". If we knew what others secretly think about us, we would have far more conflicts.

The importance of unobserved opinion formation is demonstrated by the crucial role of anonymous votes in democracies. Would we only adjust ourselves to expectations of others, many new ideas would not emerge or spread. Social diversity would decrease, and thus the ability of our society to adapt. Innovation requires the protection of minorities and new ideas. It is an engine of our economy. Social diversity also promotes happiness, social well-being, and the ability of our society to recover from shocks ("social resilience").

Social diversity must be protected just as biodiversity. Today, however, the Internet recommends us books, music, movies and even suggests how to think about politics and other people. This undermines the "wisdom of crowds" and collective intelligence. Why should a company decide what is good for us? Why can't we determine the recommendation algorithms ourselves? Why don't we get access to relevant data?

*9. An Alternative Vision of the Information Age*

In an increasingly unstable world, surveillance, combined with the manipulation or suppression of undesired behaviors, is not a sustainable solution. But is there an alternative to the omniscient almighty state that matches our ethical values? An alternative that can create cultural and economic prosperity? Yes, indeed!

Our society and economy are currently undergoing a fundamental transformation. Global networking creates increasing complexity and instability that cannot be properly managed by planning, optimization and top-down control. A flexible adaptation to local needs works better for complex, variable systems. This means that managing complexity requires a stronger bottom-up component.

In the economy and the organization of the Internet, decentralized self-organization principles have always played a big role. Now they have also spread to intelligent energy networks ("smart grids") and traffic control. One day, societal decision-making and economic production processes will also be run in a more participatory way to better manage the increase in complexity. It seems the natural course of history. A growing desire of citizens to participate in social, political and economic affairs is already found in many parts of the world.

*10. The Democratic, Participatory Market Society*

In connection with a participatory economy, one often speaks of "prosumers", i.e. co-producing consumers. Advanced collaboration platforms will allow anyone to set up projects with others to create their own products, for example with 3D printers. Thus, classical companies and political parties and institutions might increasingly be replaced by project-based initiatives – a form of organization that I would like to call "democratic, participatory market society".

To ensure that the participatory market society will work well and create jobs on a large scale, the right decisions will have to be taken. For example, it seems essential that the information systems of the future will be open, transparent and participatory. This requires us to create a participatory information and innovation ecosystem, i.e. to make large amounts of data accessible to everyone.

*11. The Benefit of Opening Data to All*

The great advantage of information is that it is (re)producible in a cheap and almost unlimited way, so that the eternal struggle for limited resources might be overcome. It is important that we take advantage of this and open the door to an age of creativity rather than limiting access to information, thereby creating artificial scarcity again. Today, many companies collect data, but lack access to other important data. The situation is as if everyone owned a few words of our language, but had to pay for the use of all the other words. It is pretty clear that, under such conditions, we could not fully capitalize on our communicative potentials.

To overcome this dissatisfactory data exchange situation and achieve "digital literacy", one could decide to open up data to all. Remember that in the past most countries decided to turn the privilege of reading and writing into a public good by establishing public schools. This step boosted the development of modern societies. Similarly, "Open Data" could boost the development of the information society, but the producers of data must be adequately compensated.

*12. A New Paradigm to Manage Complexity*

Access to data is essential for the successful management of complex dynamical systems, as it requires three elements: (i) proper systems design to support predictability and controllability, (ii) probabilistic short-term forecasts of the system dynamics, which need plenty of reliable real-time data, and (iii) suitable adaptive mechanisms ("feedback loops") that support the desired system behavior.

Managing complexity should build on the natural tendency of complex dynamical systems to self-organize. To enable self-organization, it is crucial to find the right institutional settings and suitable 'rules of the game", while avoiding too much top down control. Then, complex systems can essentially regulate themselves.

One must be aware, however, that complex systems often behave in counterintuitive ways. Hence, it is easy to choose the wrong rules, thereby ending up with suboptimal results, unwanted side effects, or unstable system behaviors that can lead to man-made disasters. The financial system, which went out of control, might serve as a warning. These problems have traditionally been managed by top-down regulation, which is usually inefficient and expensive.

*13. Loss of Control due to a Wrong Way of Thinking*

Whether a system can be adequately managed or is self-organizing in the way we want is a matter of systems design. If the system is designed in the wrong way, then it will get out of control sooner or later, even if all actors involved are highly trained, well equipped and highly motivated to do the right things. "Phantom traffic jams" and "crowd disasters" are examples of unwanted situations occurring despite all efforts to prevent them. Likewise, financial crises, conflicts and wars can be unintended consequences of systemic instabilities. Even today, we are still not immune to them.

Therefore, we need a deeper understanding of our techno-socio-economic-ecological systems and their interdependencies. Appropriate institutions and rules for our highly networked world must still be found. The information age is revolutionizing our economy and society in a dramatic way. If we do not pay sufficient attention to these developments, we will suffer the fate of driving a car too fast on a foggy day.

*14. Decisions Needed to Use Opportunities and Avoid Risks*

To meet the challenges and benefit from the great opportunities of the 21st century, a Global Systems Science needs to be established in order to fill the current knowledge gaps. It aims to generate new insights allowing politics, economy and society to take better informed, more successful decisions. This could help us to use the chances of the information age and minimize its risks. We must be aware that everything is possible – ranging from a Big Brother society to a participatory economy and society. The choice is ours!

*Further Reading*

Difficulty to anonymize data:

1. Researchers reverse Netflix anonymization, see http://www.securityfocus.com/news/11497
2. Unique in the crowd: The privacy bounds of human mobility, see http://www.nature.com/srep/2013/130325/srep01376/full/srep01376.html

Danger of surveillance society

3. Big data is opening doors, but maybe too many http://www.nytimes.com/2013/03/24/technology/big-data-and-a-renewed-debate-over-privacy.html?ref=stevelohr&_r=2&
4. Future planet – future of surveillance, see http://www.international.to/index.php?option=com_content&view=category&id=94&layout=blog&Itemid=104
5. CIA and FBI strategies to mine personal data, see http://www.businessinsider.com/cia-presentation-on-big-data-2013-3?op=1, http://gigaom.com/2013/03/20/even-the-cia-is-struggling-to-deal-with-the-volume-of-real-time-social-data/2/, http://www.slate.com/blogs/future_tense/2013/03/26/andrew_weissmann_fbi_wants_real_time_gmail_dropbox_spying_power.html

New deal on data, how to consider consumer interests:

6. US Consumer Privacy Bill of Rights, see http://www.whitehouse.gov/sites/default/files/privacy-final.pdf
7. Personal data: The emergence of a new asset class, see http://www.weforum.org/reports/personal-data-emergence-new-asset-class
8. HP software allowing personalized advertisement without revealing personal data to companies, contact: Prof. Dr. Bernardo Huberman <huberman@hpl.hp.com>

FuturICT initiative (http://www.futurict.eu):

9. FuturICT – The road towards ethical ICT, see http://link.springer.com/article/10.1140%2Fepjst%2Fe2012-01691-2#page-1
10. From social data mining to forecasting socio-economic crises, see http://epjst.epj.org/index.php?option=com_article&access=standard&Itemid=129&url=/articles/epjst/abs/2011/04/epjst195002/epjst195002.html

*About the Author*

Dirk Helbing is Professor of Sociology, in particular of Modeling and Simulation, and member of the Computer Science Department at ETH Zurich. He earned a PhD in physics and was Managing Director of the Institute of

Transport & Economics at Dresden University of Technology in Germany. He is internationally known for his work on pedestrian crowds, vehicle traffic, and agent-based models of social systems. Furthermore, he coordinates the FuturICT Initiative (http://www.futurict.eu), which focuses on the understanding of techno-socio-economic systems, using Big Data. His work is documented by hundreds of scientific articles, keynote lectures and media reports worldwide. Helbing is elected member of the World Economic Forum's Global Agenda Council on Complex Systems and of the German Academy of Sciences "Leopoldina". He is also Chairman of the Physics of Socio-Economic Systems Division of the German Physical Society and co-founder of ETH Zurich's Risk Center.